\documentclass[aps,prd,showpacs,eqsecnum,twocolumn]{revtex4}
\usepackage{amsmath,amssymb}
\usepackage{graphicx}
\begin{document}

\title {Gravitational Waves from Chaotic Dynamical System}
\author{Kenta Kiuchi$^{1}$
\footnote{e-mail address:kiuchi@gravity.phys.waseda.ac.jp}}
\author{Kei-ichi Maeda$^{1,2,3}$
\footnote{e-mail address:maeda@gravity.phys.waseda.ac.jp}}
\address{$^{1}$Department of Physics, Waseda University, 3-4-1 Okubo,
Shinjuku-ku, Tokyo 169-8555, Japan~}
\address{$^2$ Advanced Research Institute for Science and Engineering,
Waseda  University, Shinjuku, Tokyo 169-8555, Japan~}
\address{$^3$ Waseda Institute for Astrophysics, Waseda University,
Shinjuku,  Tokyo 169-8555, Japan~}

\date{\today}

\begin{abstract}
To investigate how chaos affects gravitational waves, we study the
gravitational waves from a spinning test particle moving around a Kerr
black hole, which is a typical chaotic system. 
To compare the result with those in non-chaotic dynamical system,
we also analyze a spinless test particle, which orbit can be
complicated in the Kerr back ground although the system is integrable. 
 We estimate
the emitted gravitational waves by the multipole
expansion of a gravitational field. We
find a striking difference in the energy spectra of the gravitational
waves. 
The spectrum for a chaotic orbit of a spinning particle,
contains various frequencies, while some characteristic
frequencies appear in the case of a spinless particle.
\end{abstract}


\maketitle

\section{Introduction}
 
The current ground-base gravitational wave detectors such as 
LIGO, TAMA300, GEO600 and
VIRGO~~\cite{LIGO,TAMA,GEO,VIRGO} are now starting
the ``science run" and the laser space antenna for gravitational
waves(LISA)~\cite{LISA} may operate  in the near future.
Gravitational waves will  bring us various new information about
relativistic astrophysical objects. If we will detect gravitational waves
and compare them with theoretical templates, we may be able to determine a
variety of astrophysical parameters of the sources such as their
direction, distance, masses, spin, and so on. 
The direct observation of
gravitational waves could resolve strong-gravitational phenomena such as
a black hole formation.
Furthermore, we may be able not only to verify the theory of
gravity but also to find new information or to recover new physics at 
high density or high energy region.
Therefore we
need to make theoretical templates of gravitational waves from various
astrophysical objects and phenomena to extract useful information from
gravitational waves.

There are a lot of astrophysical objects and phenomena  as gravitational
wave sources, e.g., a closed binary system or a supernova explosion.
Many attempts to study their templates have been done so far.
In this paper, we particularly focus on a
gravitational waves form a chaotic dynamical system. 
It is motivated as follows. Chaos appears  
 universally in nature  and it is
expected to explain various nonlinear phenomena. 
Because chaos
indicates unpredictability of motion of objects, it comes into a question
whether or not that it is possible to extract useful information from
a chaotic system. Many studies have been made on
chaos in mechanics and astrophysics since the research on the three-body
problem by Poncar\'{e}. It is also  important to study chaos in
general relativity because the Einstein equations are nonlinear. Many
authors have reported chaos in general
relativity
~\cite{Hobill,Barrow,Conto,Dettmann,Yust,Karas,Varvoglis,Bombelli,Moeckel,
Sota,Suzuki,Suzuki2,Suzuki3,Mich,Seme,Levin,Schnittman,Cornish}. As for a
realistic astrophysical object,  there has been a discussion whether or not chaos
occurs in a compact binary system~\cite{Levin,Schnittman,Cornish}. It seems that
there is not enough time in a compact binary for chaos to occur even in a
non-integrable system. Then we may not have so far any concrete example of
general  relativistic chaos for a realistic situation.
However, we believe that it is necessary to
study chaos in general relativity. Because  chaos will
occur in a realistic strong gravitational field described by general
relativity.

In this paper, we study some observational feature of chaos in general 
relativistic  dynamical system. We focus on gravitational waves from a chaotic
dynamical system. As a model of a chaotic dynamical system, we adopt a spinning
test particle  around a rotating black hole. It mimics a model of
compact objects orbiting around a super massive black hole. Recent
observations support the existence of such massive black holes and LISA
will be sensitive to the gravitational wave from such systems. As for a
background spacetime, we assume the Kerr metric since such a supermassive
black hole is usually rotating. The previous
works~\cite{Suzuki,Seme} revealed that orbits of a spinning test
particle around a black hole can be chaotic due to the
spin-orbit coupling. We evaluate the gravitational waves emitted from this
system by using multipole expansion of a gravitational
field.~\cite{Landau,Mathews,Kip}.

Carter showed~\cite{Cart} that the equations  of motion for  a 
spinless particle are integrable. After his work, Johnston 
discovered that motion for some initial parameters,
the particle motion can
be very complicated
as if chaos occurred, although the system is integrable~\cite{John}.
Then we also analyze such non-chaotic but complicated orbital motion
to clarify whether or not chaos is essential in our results.
 We compare the wave forms and the energy
spectra of the gravitational waves from both chaotic and nonchaotic
systems and analyze the effects of chaos on gravitational waves.

This paper is organized as follows. In Sec. II we shall briefly  review
the basic equations of both a spinless particle and a spinning particle
in relativistic spacetime, a supplementary condition, and some constants
of motion. Specifying the background spacetime to be the Kerr metric,
we show our results in Sec. III. Summary and
some remarks follow in Sec. IV.
\\
Throughout this paper we use units $c=G=1$. Our notation including  the
signature of the metric follows that of Misner Thorne and
Wheeler(MTW)~\cite{MTW}.


\section{Basic equations for a test particle}
First we explain our system more precisely.
\subsection{Background spacetime}
We consider the Kerr metric as a background spacetime.  In the
Boyer-Lindquest coordinates, it is given by
\begin{align}
ds^2 = &- \Big(1-\frac{2Mr}{\Sigma}\Big)dt^2 - 
\frac{4Mar\sin^2\theta}{\Sigma}dtd\phi \nonumber\\
+&\frac{\Sigma}{\Delta}dr^2 + \Sigma d\theta^2 +
\sin^2\theta\Big(r^2+a^2+\frac{2Ma^2r\sin^2\theta}{\Sigma}\Big)d\phi^2,
\end{align}
where
\begin{align}
\Sigma &= r^2 + a^2\cos^2\theta, \\
\Delta &= r^2 - 2Mr + a^2,
\end{align}
and $M$ and $a$ are the mass of a black 
hole and its angular momentum, respectively.

\subsection{A spinless test particle}
\label{Sec:eq of spinless}
In this subsection, we summarize the equations of motion for a spinless 
test particle. It is well known that a spinless test particle moves
along geodesic
\begin{align}
\frac{d^2 x^\mu}{d\tau^2} + \Gamma^\mu_{\nu\rho} \frac{dx^\nu}{d\tau} 
\frac{dx^\rho}{d\tau} = 0,\label{geo1}
\end{align}
where $\tau$ is an affine parameter of the orbit. In a stationary 
axisymmetric spacetime, a spinless particle has
two constants of motion; the Energy $E$ and the $z$-component of the
angular momentum $L_z$. The particle's rest mass $\mu$ is also constant.
Carter discovered third constant of motion and showed that the system
is integrable~\cite{Cart}.
Eq. (\ref{geo1}) is reduced to a set of the
differential equations as
\begin{align}
&\Sigma\frac{d\theta}{d\tau} = \pm\sqrt{\Theta} \label{nonchaos1},\\
&\Sigma\frac{dr}{d\tau}= \pm\sqrt{R} \label{nonchaos2}, \\
&\Sigma\frac{d\phi}{d\tau} = - \Big(aE-\frac{L_z}{\sin^2\theta}\Big) + 
\frac{a}{\Delta}P \label{nonchaos3},\\ &\Sigma\frac{dt}{d\tau} = -
a(aE\sin^2\theta - L_z) + \frac{r^2+a^2}{\Delta}P\label{nonchaos4},
\end{align}
where
\begin{align}
\Theta(\theta) &= C - \cos^2\theta\Big\{a^2(1-E^2)
+\frac{L_z^2}{\sin^2\theta}\Big\}, \\ P(r) &= E(r^2+a^2) - aL_z, \\
R(r) &= P^2 - \Delta\{r^2+(L_z-aE)^2 +C\},
\end{align}
and $C$ is a constant discovered by Carter and called the Carter constant.
Note that 
because of the presence of the Carter constant, the orbits of
a particle will never be chaotic.

\subsection{A spinning test particle}\label{Sec:eq of spin}
The equations of motion for a spinning test particle in a  relativistic
spacetime were first derived by Papapetrou~\cite{Papa} and then
reformulated by Dixon~\cite{Dix}. They used the pole-dipole approximation,
in which multipole moments of an object higher than a mass
monopole and a spin dipole are ignored in the limit of a point particle.
The set of equations is given as 
\begin{align}
&\frac{dx^\mu}{d\tau} = v^\mu, \label{basic1}\\
&\frac{Dp^\mu}{d\tau} = - \frac{1}{2}{R^\mu}_{\nu\rho\sigma}  v^\nu
S^{\rho\sigma}, \label{basic2}\\ &\frac{D S^{\mu\nu}}{d\tau} = 2 p^{[\mu}
v^{\nu]}, \label{basic3}
\end{align}
where $\tau,v^\mu,p^\mu$ and $S^{\mu\nu}$ are an affine parameter  of the
orbit, the four-velocity of a particle, the momentum, and the spin
tensor, respectively. $p^\mu$ deviates from a geodesic due to the
coupling of the Riemann tensor with the spin tensor.
We also need an additional
 condition which gives a relation between $p^\mu$ and
$v^\mu$. We adopt the condition formulated by Dixon~\cite{Dix}
\begin{align}
p_\mu S^{\mu\nu} = 0 \label{cond1}.
\end {align}
This condition consistently determines 
the cetnter of mass in the present
system. Using Eq (\ref{cond1}), we can determine the relation between
$p^\mu$  and $v^\mu$ explicitly as
\begin{align}
v^\mu = N\Big[u^\mu + \frac{1}{2\mu^2 \delta}
R_{\nu\lambda\rho\sigma}S^{\mu\nu} u^\lambda S^{\rho\sigma}\Big],
\end{align}
where 
\begin{align}
\delta = 1 + \frac{1}{4\mu^2}R_{\alpha\beta\gamma\sigma}S^{\alpha\beta} 
S^{\gamma\sigma},
\end{align}
and $N$ is a normalization constant, which is fixed by a choice of the 
affine parameter $\tau$ as we will explicitly show it later.
$u^\mu \equiv p^\mu/\mu$ is a unit vector parallel to the four-momentum 
$p^\mu$, where the mass of the particle $\mu$ is defined by 
\begin{align}
\mu^2 = - p_\mu p^\mu.\label{const1}
\end{align}
To make clear the freedom of this 
system, we have to check the conserved quantities. Regardless of the
symmetry of the background spacetime, it is easy to show that $\mu$ and
the magnitude of spin $S$ defined by
\begin{align}
S^2 \equiv \frac{1}{2}S_{\mu\nu} S^{\mu\nu},
\end{align}
are constants of motion. If a background spacetime possesses some
symmetry  described by a Killing vector $\xi^\mu$,
\begin{align}
C_\xi \equiv \xi^\mu p_\mu - \frac{1}{2}\xi_{\mu;\nu} S^{\mu\nu}
\end{align}
is also conserved~\cite{Dix}. For a spacetime with both axial  and
timelike Killing vectors such as Kerr spacetime, we have two conserved
quantities, i.e., the Energy $E$ and the $z$-component of the total
angular momentum $J_z$ of a spinning particle.
 
For numerical reasons,
we rewrite our basic equations as follows. 
Following the appendix
in~\cite{Suzuki}, we write down the equations by use of a spin vector
\begin{align}
S_\mu \equiv  - \frac{1}{2}\epsilon_{\mu\nu\rho\sigma}u^\nu 
S^{\rho\sigma}, \label{svec}
\end{align}
where $\epsilon^{\mu\nu\rho\sigma}$ is the Levi-Civita tensor.
The basic equations are now
\begin{align}
\frac{d x^\mu}{d \tau} &= v^\mu, \label{basic4}\\
\frac{D p^\mu}{d \tau} &=  \frac{1}{\mu} {R^{*\mu}}_{\nu\rho\sigma}v^\nu
S^\rho p^\sigma, \label{basic5}\\
\frac{D S^\mu}{d \tau} &= 
\frac{1}{\mu^3}p^\mu{R^{*}}_{\nu\rho\sigma\gamma}S^\nu v^\rho S^\sigma
p^\gamma,\label{basic6}
\end{align} 
where
\begin{align}
{R^{*}}_{\mu\nu\rho\sigma} \equiv \frac{1}{2} {R_{\mu\nu}}^{\alpha\beta} 
\epsilon_{\alpha\beta\rho\sigma}.
\end{align}
Eq. (\ref{cond1}) is also rewritten
\begin{align}
p_\nu S^\nu = 0. \label{cond2}
\end{align}
By this condition, the relation between $p^\mu$ and $v^\mu$ is determined 
as follow :
\begin{align} 
v^\mu = u^\mu + \frac{1}{\mu^2} {^*R^{*\mu}}_{\nu\rho\sigma}S^\nu S^\rho 
u^\sigma,
\end{align}
where
\begin{align}
{^*R^{*}}_{\mu\nu\rho\sigma} \equiv \frac{1}{2} 
\epsilon_{\mu\nu\alpha\beta} {R^{*\alpha\beta}}_{\rho\sigma}.
\end{align}
We fix the affine parameter $\tau$ using the condition  $v^\mu u_\mu =
-\delta$. This gives
$N$ as
\begin{align}
N=1 + \frac{1}{\mu^4}{^*R}^*_{\alpha\beta\mu\nu}S^\alpha p^\beta 
S^\mu p^\nu,
\end{align}
Our conserved quantities are also
rewritten as
\begin{align}
&S^2 = S_\mu S^\mu, \label{const2}\\
&C_\xi = \xi^\mu p_\mu - \frac{1}{2\mu}\xi_{\mu;\nu}
\epsilon^{\mu\nu\rho\sigma}p_\rho S_\sigma.
\label{const3}
\end{align}

\section{Numerical results}\label{Sec:Num}
In this section, we show our numerical results for the orbital motion of a
test particle, and the wave forms and the energy spectra of the emitted
gravitational waves. 
To analyze chaotic behavior of a test particle, we use the Poincar\'{e}
map and the Lyapunov exponent. To draw the Poincar\'{e} map, we adopt the
equatorial plane ($\theta =
\pi/2$) as a Poncar\'{e} section and plot the point ($r$, $v^r$)  when the
particle crosses the Poincar\'{e} section with $v^\theta < 0$. If the
motion is not chaotic, the plotted points will form a closed curve in the
two dimensional $r$-$v^r$ plane, because a regular orbit will move on a
torus in the phase space and the curve is a cross section of the torus.
If the orbit is chaotic, some of these tori will be broken and the
Poincar\'{e} map does not consist of a set of closed curves but the
points will be distributed randomly in the allowed region. From the
distribution of the points, we can judge whether or not the motion is
chaotic.

\subsection{Motion of a test particle}\label{Sec:motion of 
spinless}
\subsubsection{a spinless particle}
\begin{figure}
\includegraphics[width=5cm]{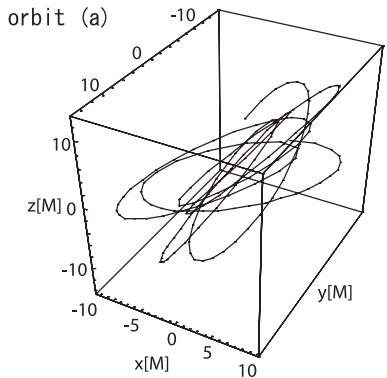}
\caption{\label{fig:nonchaos}The orbit of a spinless particle with 
$E=0.968\mu,J_z=2.0\mu M,C = 10\mu^2 M^2$ and $a = 1/\sqrt{2}$. We choose
the initial position and velocity as $(r,\theta, \phi)=(10M,\pi/2,0)$ and
$(v^r,v^\theta,v^\phi)=(0.14,0.03,0.02)$. This
 is called the orbit (a) in this paper. This orbit looks very
complicated as if chaos occurred.}
\end{figure}

\begin{figure}
\includegraphics[width=5cm]{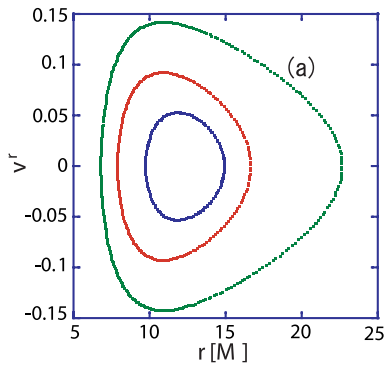}
\caption{\label{fig:pmap1}The Poincar\'{e} map of the orbit (a).  The
plot points form a closed curve in the two-dimensional $r$-$v^r$ plane.
We also plot the Poincar\'{e} map of other two orbits with the same conserved 
quantities as those of the orbit (a).
This confirms that the system is integrable.}
\end{figure}

\label{Sec:motion of 
spinless}
As we see in Sec \ref{Sec:eq of spinless}, the equations of the motion of 
a spinless particle are integrable because of the existence of the
Carter constant. That is chaos does not occur. But Johnston found that the
orbits of a spinless particle can be complicated as if chaos
occurred~\cite{John}.
Integrating numerically the equations of motion
~(\ref{nonchaos1})-(\ref{nonchaos4}), we show a typical complicated orbit
in Fig.~\ref{fig:nonchaos}. \\
 This orbit is called
the orbit (a) in this paper. The behavior of the orbit (a) looks
complicated. However, showing the Poincar\'{e} map in
Fig.~\ref{fig:pmap1}, we find a closed curve, which confirms that the
orbit (a) is not chaotic.  
\subsubsection{a spinning particle}
\begin{figure}
\includegraphics[width=7cm]{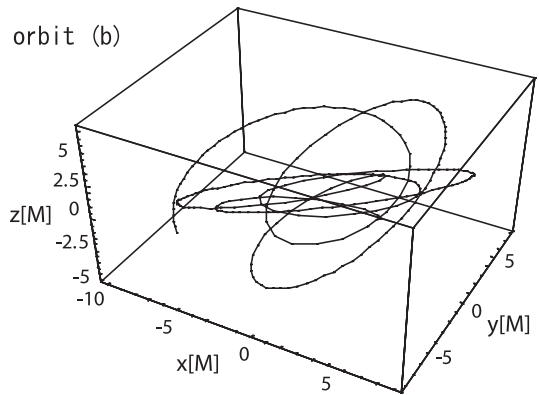}
\caption{\label{fig:chaos}The orbits of a spinning particle with  $E =
0.9328\mu,J_z = 2.8\mu M,$ and $ S = 1.0\mu M $. We set $a = 0.8M$ and 
choose the initial position and velocity as $(r,\theta, \phi)=(6.0M,\pi/2,0)$
and
$(v^r,v^\theta,v^\phi)=(0.18,0.05,0.07)$.
This is called the orbit
(b), which is.
 complicated just as the orbit (a).}
\end{figure}

\begin{figure}
\includegraphics[width=5cm]{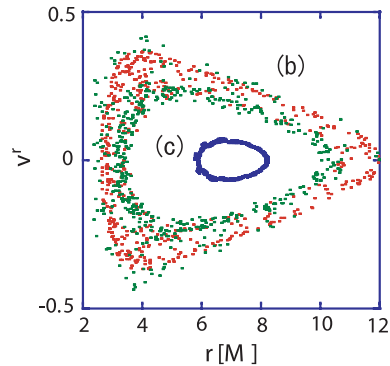}
\caption{\label{fig:pmap2}The Poincar\'{e} maps of the orbit (b) and (c).
The  plot points for the orbit (b) distribute randomly, which confirms 
that it is a chaotic system. For the orbit (c), we will discuss it later.}
\end{figure}

\begin{figure}
\includegraphics[width=7cm]{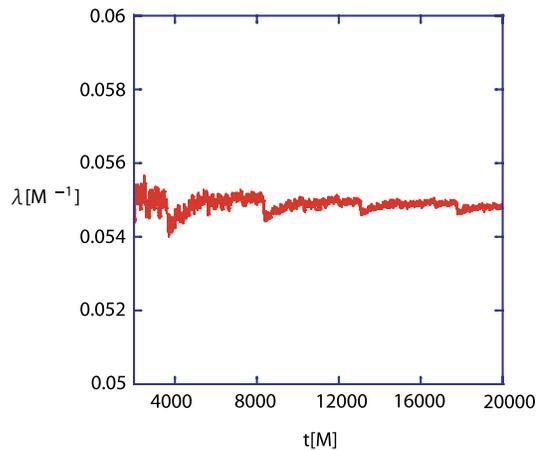}
\caption{\label{fig:Lyap}The maximal Lyapunov exponent $\lambda$ for the
orbit (b).  The typical time scale for chaos is given by
$\lambda^{-1}\sim 18 M$, while the average of orbital period is $74.3M$.}
\end{figure}

\label{Sec:motion  of
spinning}
In the previous work~\cite{Suzuki2,Suzuki3}, the parameter region
for a spinning test particle in which a particle will show chaotic
behavior was discussed. 
We adopt such parameters, e.g. the
energy $E$, the angular momentum $J_z$, and the spin parameter $S$ 
of a test
particle. 
Using the
Bulirsch-Store method~\cite{numerical}, we integrate the equations of
motion Eqs.~(\ref{basic4})-(\ref{basic6}) and use the constraint
equations
Eqs.~(\ref{const1}),(\ref{cond2}),(\ref{const2}),(\ref{const3}) to check
the accuracy of our numerical integration. We find that the relative
errors are smaller than $10^{-11}$ for each constraint. 
We show the
orbits of the spinning particle in Fig.~\ref{fig:chaos} and call it the
orbit (b). 
Comparing Fig.~\ref{fig:nonchaos} with Fig.~\ref{fig:chaos}, we cannot
distinguish two orbits. However, the difference between chaotic
orbit and nonchaotic one will be apparent when we draw the Poincar\'{e}
maps. In Fig.~\ref{fig:pmap2}, we show the Poincar\'{e} map of the
orbit (b). 
The plot points distribute randomly in Fig.~\ref{fig:pmap2}. We
also calculate  the Lyapunov exponent $\lambda$ to evaluate
the strength of chaos. The result is depicted in Fig. \ref{fig:Lyap}.
As shown in  Fig. \ref{fig:Lyap}, the Lyapunov exponent $\lambda$ is
positive, which means that the orbit (b) is chaotic.
The typical time scale for chaos is given by
$\lambda^{-1}\sim 18 M$.
Since the average of orbital period is $74.3M$,
the orbit (b) becomes chaotic just after a few revolutions around a black
hole.

\subsection{Gravitational wave forms and energy spectra}\label{Sec:GW}
Based on the previous calculation of the orbits, we show the wave forms
and its energy spectra and  analyze an effect of chaos on the
emitted gravitational waves.
 To estimate the gravitational waves, we use the multipole
expansion of gravitational field~\cite{Mathews,Kip}. 
Here we briefly summarize the  multipole
expansion method. 
According to ~\cite{Mathews,Kip}, the gravitational
waves are represented in TT gauge as
\begin{align}
h^{TT}_{ij} = \sum^\infty_{l=2}\sum^l_{m=-l} \frac{1}{r}
\Big(\frac{d^l}{dt^l} I^{lm}(t-r) f_{ij}^{lm} + \frac{d^l}{dt^l}
S^{lm}(t-r) d_{ij}^{lm} \Big),
\end{align}
where
\begin{align}
&f_{lm} = \frac{r^2}{\sqrt{4nl(l+1)}}\left(\begin{array}{ccc}
0&0&0 \\
0&W_{lm}&X_{lm} \\
0&*& -\sin^2\theta W_{lm}
\end{array}\right),
\nonumber \\
&d_{lm} = \frac{r^2}{\sqrt{4nl(l+1)}}\left(\begin{array}{ccc}
0&0&0 \\
0&-X_{lm}/\sin\theta&\sin\theta W_{lm} \\
0&*& \sin\theta X_{lm}
\end{array}\right),
\end{align}
with 
\begin{align}
&n = \frac{l(l+2)}{2} -1,\\
&X_{lm} = 2\frac{\partial}{\partial \phi}\Big(\frac{\partial}{\partial 
\theta}-\cot\theta \Big) Y_{lm},\nonumber
\\ &W_{lm} = \Big(
\frac{\partial^2}{\partial \theta^2} - \cot\theta
\frac{\partial}{\partial \theta} -
\frac{1}{\sin^2\theta}\frac{\partial^2}{\partial\phi^2}\Big)Y_{lm} \,.
\end{align}
($r$, $\theta$) is a position of an observer. $d_{lm}$ and $f_{lm}$ are an
odd-parity   and  even-parity mode, respectively. $I^{lm}$ is
a multipole of the mass distribution and $S^{lm}$ is that of the angular
momentum. If a gravitational wave source moves slowly, we can write down
$I^{lm}$ and
$S^{lm}$ as
\begin{align}
I^{lm} &= \frac{16\pi}{(2l+1)!!}\Big(\frac{(l+1)(l+2)}{2(l-1)l}
\Big)^{\frac{1}{2}}\int\tau_{00} Y^*_{lm} r^l d^3x,\\ 
S^{lm} &= -
\frac{32\pi}{(2l+1)!!}\Big(\frac{(l+2)(2l+1)}{2(l-1)(l+1)}
\Big)^{\frac{1}{2}}
\nonumber \\
&~~~\times \int
\epsilon_{jpq} x_p(-\tau_{0q})Y^{l-1,lm*}_j r^{l-1} d^3x,
\end{align}  
where $Y^{l-1,lm*}_j$ is called the pure orbital harmonic function
\begin{align}
&Y^{l',lm}_j(\theta,\phi) = 
\sum^{l'}_{m'=-l'}\sum^1_{m''=-1}(1l'm''m'|lm)\xi_i^{m''} Y^{l'm'},\\
&\xi^0_i \equiv (\textbf{e}_z)_i,~~~\xi^{\pm}_i \equiv
\mp\frac{1}{\sqrt{2}}(\textbf{e}_x \pm i \textbf{e}_y)_i\,.
\end{align}
Giving the information of a particle orbit, we can evaluate 
$h_{ij}^{TT}$ for each $l$.
The gravitational waves from a chaotic orbit of a spinning
particle may contain higher multipole moments ($l>2$)
than quadrupole ($l=2$).
This is the reason why we analyze the gravitational waves not only $l=2$
but also 
$l = 3$. Note that because the orbit considered here is very
relativistic, the multipole expansion may not be valid and then
it will provide only a qualitative feature.

\begin{figure}
\includegraphics[width=8cm]{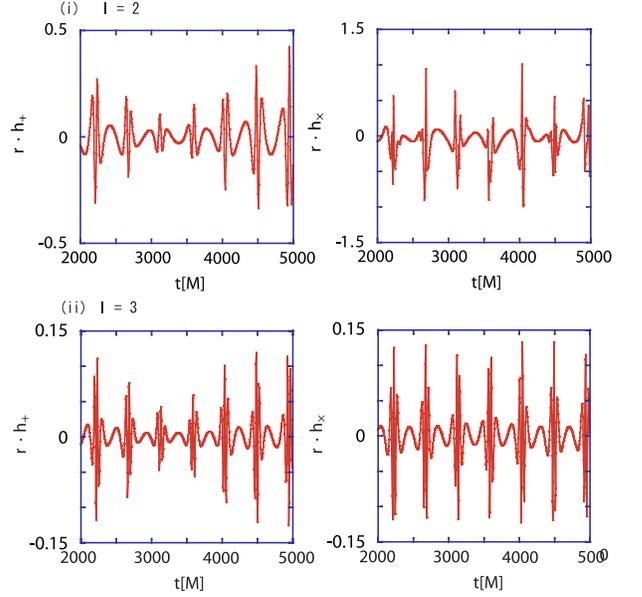}
\caption{\label{fig:wave_a}
The gravitational wave forms of the $+$- and 
the $\times$-modes for the orbit (a).  The observer is on the equatorial
plane. Figs.~(i) and (ii) show the wave forms of $l$= 2 and 3,
respectively.}
\end{figure}

\begin{figure}
\includegraphics[width=8cm]{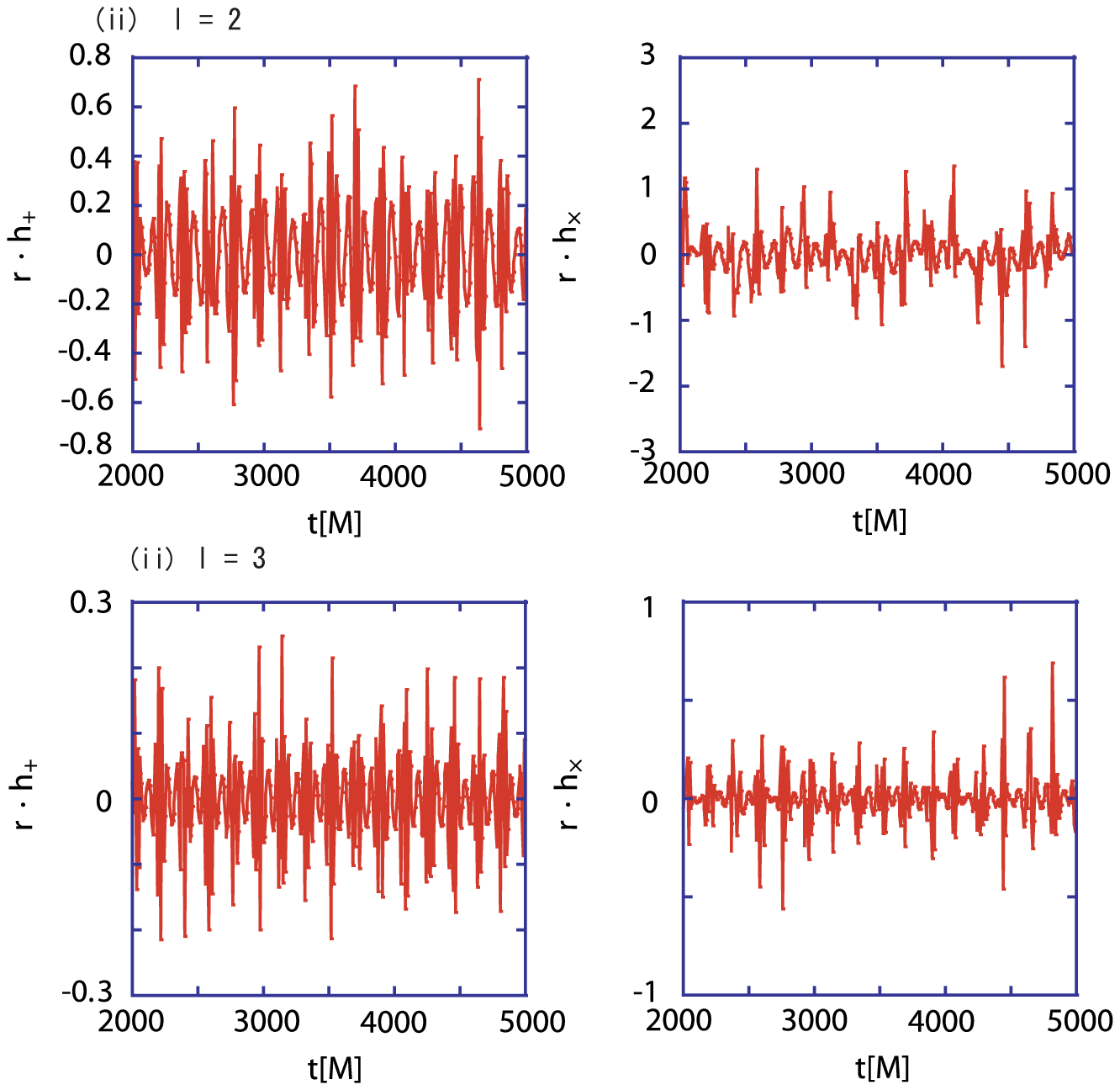}
\caption{\label{fig:wave_b}
The gravitational wave forms of the $+$- and 
the $\times$-modes for the orbit (b). The observer is on the equatorial
plane. Figs.~(i) and (ii) show the wave forms of $l$= 2 and 3,
respectively.}
\end{figure}

In Fig.~\ref{fig:wave_a} and Fig.~\ref{fig:wave_b}, 
we show the gravitational wave forms for the orbits (a) and (b),
respectively.
\begin{figure}
\includegraphics[width=8cm]{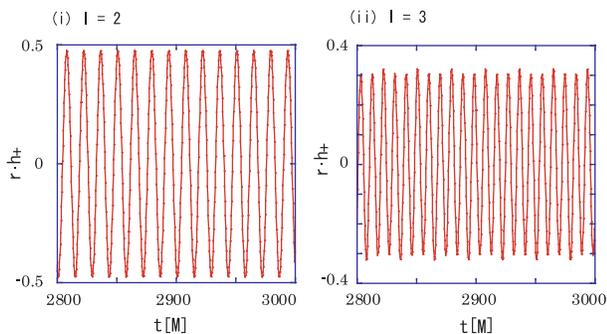}
\caption{\label{fig:cir_wave}
The wave forms for a circular orbit at
$r=2.45M$.}
\end{figure} 
These figures reveal  two important points. 
One is that  in the wave forms, there looks some difference between two
orbits (a) and (b), but it may be difficult to distinguish which one is
chaotic. Just as reference, we depict the wave form for a circular orbit
in Fig. \ref{fig:cir_wave},
which is quite regular.
 The second points is that
the amplitudes of the octopole wave ($l=3$) are smaller than those of the
quadrupole wave ($l=2$). Comparing the peak values, we find that the
ratio of
 the amplitude of $l=3$ to that of 
$l=2$  is about 30 percents.

\begin{figure}
\includegraphics[width=6cm]{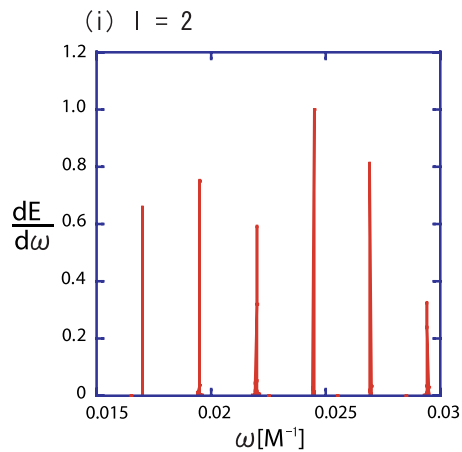}
\caption{\label{fig:spnonchaos}
A energy spectrum of the gravitational 
wave from the orbit (a). 
Each value is normalized by
the maximum peak value.}
\end{figure}

In Fig.~\ref{fig:spnonchaos},
we show the energy spectrum of the quadrupole gravitational wave
($l=2$) from the orbit (a).
\begin{figure}
\includegraphics[width=7cm]{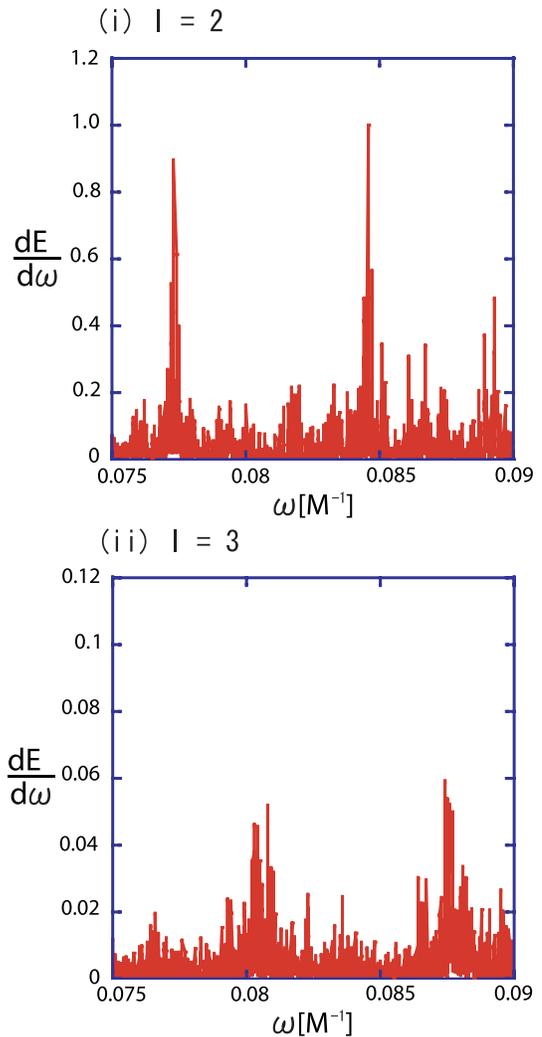}
\caption{\label{fig:spchaos}The energy spectra of the gravitational  wave
($l=2,3$) from the spinning particle. (i)-(ii)The energy spectrum of the
gravitational wave ($l=2,3$). The values in these case are normalized by the
peak value of the spectrum $l=2$. }
\end{figure}
In Fig.~\ref{fig:spchaos}, we show the energy spectrum of the 
gravitational wave($l=2,3$) from the spinning particle.
From Fig.~\ref{fig:spnonchaos} and Fig.~\ref{fig:spchaos}, we find that 
there is a clear difference in the energy spectra of the gravitational
waves between the orbit (a) and (b). 

\begin{figure}
\includegraphics[width=5cm]{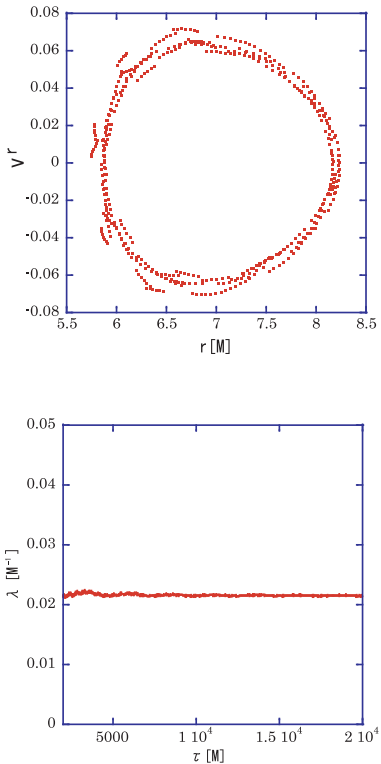}
\caption{\label{fig:spin-reg_Poincare}
(i) The Poincar\'{e} map of the orbit
(c).  We also plot the Lyapunov exponent (ii), which value is smaller than
that of the orbit (b) but is still positive ($\lambda \sim 0.02 M$).}
\end{figure}

\begin{figure}
\includegraphics[width=8cm]{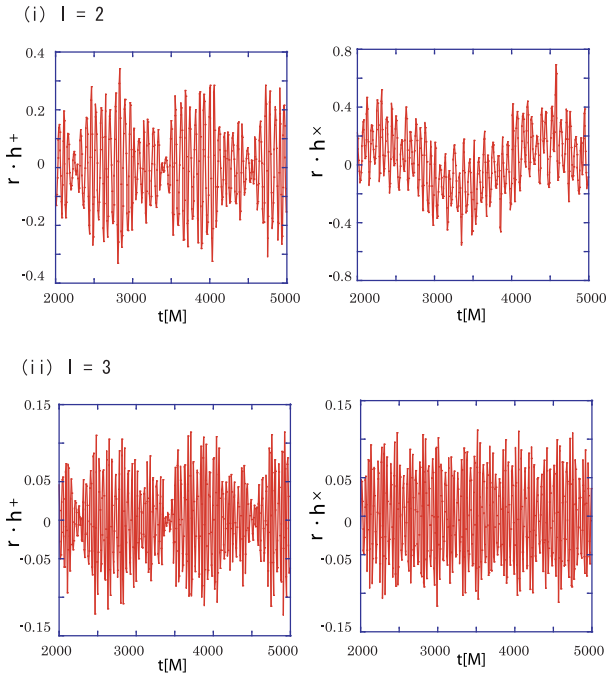}
\caption{\label{fig:spin-reg_wave}
The gravitational wave forms of the $+$- and 
the $\times$-modes for the orbit (c).  The observer is on the equatorial
plane. Figs.~(i) and (ii) show the wave forms of $l$= 2 and 3,
respectively.}
\end{figure}

\begin{figure}
\includegraphics[width=6cm]{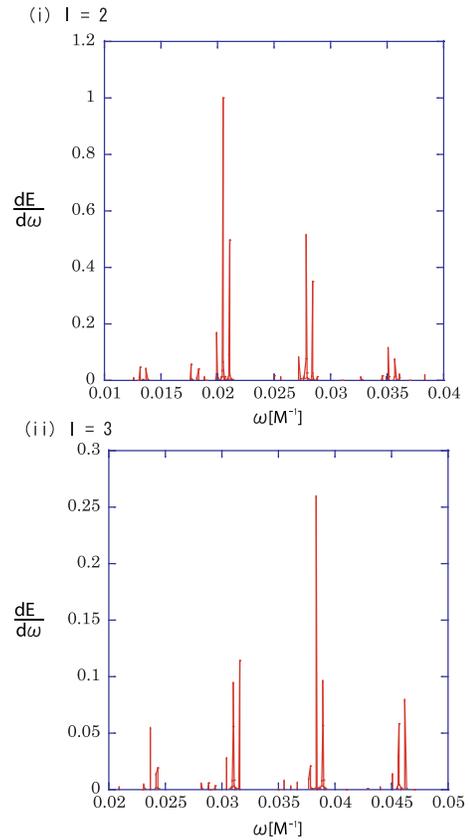}
\caption{\label{fig:spin-reg_spectrum}
A energy spectrum of the gravitational 
wave from the orbit (c). This is similar to the orbit (a).}
\end{figure}

How about the orbit of a spinning particle which looks regular?
In order to understand a role of chaotic behavior, we
also analyze the orbit (c), which Poincar\'{e} map is nearly a closed
circle (the inner circle map in Fig. \ref{fig:pmap2}).
Although this map looks closed in Fig. \ref{fig:pmap2}, we find that it is
not really closed when we enlarge it (see Fig.
\ref{fig:spin-reg_Poincare}).
 We show the wave forms for the orbit (c) in
Fig.
\ref{fig:spin-reg_wave}.
We also present the spectrum in Fig. \ref{fig:spin-reg_spectrum}.


From the figures, we may conclude that 
the spectrum of gravitational waves from nonchaotic orbits contains only
discrete characteristic frequencies. 
On the other
hand, the spectrum for the chaotic orbits
contains the various frequencies. It seems to be a continuous spectrum
with finite widths. 
The ratio of the octopole wave energy for the orbit
(b) to that of the quadrupole wave is about 12.4 percents. Therefore the
higher pole moment than the quadrupole moment does not contribute so much
to the emitted gravitational waves, although the orbit is chaotic.

\section{Summary and Discussions}
In this paper, we study the  gravitational waves emitted from a chaotic
dynamical system.  As a concrete example, we analyze the motion
of a test particle going around the Kerr black hole.
We confirm that the orbit of a spinning particle can be chaotic.
Using the multipole
expansion method of a gravitational field, we evaluate the gravitational
waves from a chaotic and a non-chaotic orbit in order to analyze the
effect of chaos on the emitted gravitational waves. As the results, 
there are not so much difference for  the
gravitational wave forms. However,  the energy spectra of
the gravitational waves show a clear difference.
For a chaotic orbit, we find a continuous energy spectrum with several
peaks.
While, in the case of nonchaotic orbit, the spectrum contains the
discrete  characteristic frequencies. We also find that the higher pole
moments than quadrupole moment of the system do not contribute so much
to the emitted gravitational waves even for a chaotic orbit.

When
the gravitational waves are detected and the energy spectrum
is determined by observation, not only  the astrophysical
parameters ,e.g., the mass, the angular momentum, and the spin are
determined but also some fundamental physics such as relativistic
nonlinear dynamics
could be discussed.

As a future work, we have to study the gravitational waves in the
 dynamical system in a full relativistic approach. 
For the first step, using the Newmann-Penrose
formalism~\cite{NP}, we extend the present evaluation (the multipole
expansion) to a full relativistic perturbation analysis.

\acknowledgments

We would like to thank H. Sotani and S. Yamada for useful discussions. 
This work was partially supported by the Grant-in-Aid for
Scientific Research  Fund of the MEXT
 (No. 14540281) and by the Waseda University Grant for
Special Research Projects and  for The 21st Century 
COE Program (Holistic Research and Education Center for Physics 
Self-organization Systems) at Waseda University.


\end{document}